\documentclass[12pt]{article}
\usepackage{epsfig}
\newcommand{\abstracts}[1]{{
\centering{\begin{minipage}{12.5truecm}
\normalsize\baselineskip=15pt
\centerline{\footnotesize ABSTRACT}\vspace*{0.3cm}
\parindent=20pt #1
\end{minipage}}\par}}
\newcommand{\la}{\langle}
\newcommand{\ra}{\rangle}

\newcommand{\cD}{{\cal D}}
\newcommand{\cZ}{{\cal Z}}

\newcommand{\Z}{{Z \!\!\! Z}}
\newcommand{\beqn}{\begin{eqnarray}}
\newcommand{\eeqn}{\end{eqnarray}}
\newcommand{\eq}[1]{(\ref{#1})}
\newcommand{\dd}{\mbox{d}}

\begin{document}

~
\vspace{-1cm}
\begin{flushright}
{\large 
Leipzig NTZ 6/2001\\
LU-ITP 2001/011\\
UNITU--THEP--15/2001\\}
\vspace{0.3cm}
{\sl May 18, 2001}
\end{flushright}

\begin{center}

{\baselineskip=16pt
{\Large \bf A lattice study of $3D$ compact QED \\
at finite temperature} 

\vspace{1cm}

{\large
M.~N.~Chernodub\footnote{maxim@heron.itep.ru}$^{\!,a}$,
E.-M.~Ilgenfritz\footnote{ilgenfri@alpha1.tphys.physik.uni-tuebingen.de}$
^{\!,b}$
and A.~Schiller\footnote{Arwed.Schiller@itp.uni-leipzig.de}$^{\!,c}$}\\

\vspace{.5cm}
{ \it

$^a$ ITEP, B.Cheremushkinskaya 25, Moscow, 117259, Russia

\vspace{0.3cm}

$^b$ Institut f\"ur Theoretische Physik, Universit\"at T\"ubingen,
D-72076 T\"ubingen, Germany

\vspace{0.3cm}

$^c$ Institut f\"ur Theoretische Physik and NTZ,
Universit\"at  Leipzig, D-04109 Leipzig, Germany

}}
\end{center}
\vspace{5mm}

\abstracts{
We study the deconfinement phase transition and monopole properties 
in the finite temperature $3D$ compact Abelian gauge model on the lattice. 
We predict the critical coupling as function of the lattice size in a 
simplified model to describe 
monopole binding. We demonstrate numerically that the monopoles are 
sensitive to the transition. In the deconfinement phase the monopoles appear in 
the form of a  dilute gas of magnetic dipoles. In the confinement phase both 
monopole density and string tension differ from semiclassical estimates 
if monopole binding is neglected. However, the analysis of the monopole 
clusters 
shows that the relation between the string tension and the  density of 
monopoles in 
charged clusters is in reasonable agreement with those predictions. 
We study the cluster structure of the vacuum in both phases of the model.}

\baselineskip=14pt
%%% For alphabetic footnotes indices in text  %%%%
\setcounter{footnote}{0}
\renewcommand{\thefootnote}{\alph{footnote}}
%%%%%%%%%%%%%%%%%%%%%%%%%%%%%%%%%%%%%%%%%%%%%%%%%%

\section{Introduction}

Compact Abelian gauge theory in three Euclidean dimensions is a case 
where permanent confinement is proven and qualitatively understood
~\cite{Polyakov,Goepfert}. In order to gain some experience for
more realistic theories, it is interesting to study how 
this mechanism ceases to work under special conditions. High temperature
is such a case.  In this paper we are going to revisit the finite temperature
deconfining phase transition. We will emphasize the aspect of monopole binding
which explains the breakdown of confinement. In a companion paper, we extend 
these studies to the case of non--vanishing external fields.

Compact QED theory possesses Abelian monopoles as topological defects appearing
due to the compactness of the gauge group. Considering the $3D$ theory as the
static limit of a $4D$ theory, the monopoles are just {\it magnetic} monopoles 
at rest, and the components of field strength are {\it magnetic}.  
In a three dimensional theory, the monopoles are instanton--like objects: 
instead of tracing world lines (as they do in $4D$) they occupy points.  
The plasma of monopoles and anti--monopoles can explain the permanent 
confinement of oppositely charged electric test charges~\cite{Polyakov} in 
bound states, kept together by a linear potential.
In the language of magnetostatics, confinement appears due to the screening of
the magnetic field induced by the electric current circulating along the Wilson
loop. Monopoles and anti--monopoles form a polarized sheet of finite thickness
(``string'')
along the minimal surface spanned by the Wilson loop. 
The formation of the string (observed in the lattice simulations in 
Ref.~\cite{Sterling})
leads, for
non--vanishing electric current, to an excess of the free energy proportional
to the area. 

At finite temperature the phase structure becomes non--trivial. What we have 
in mind, is compactifying $3D  \rightarrow (2+1)D$ in the 
``temporal'' (third or $z$) direction. The con\-fi\-ne\-me\-nt\--deconfinement
phase transition was studied on the lattice  both analytically~\cite{Parga} and
numerically~\cite{Coddington}. According to the Svetitsky--Yaffe universality 
arguments~\cite{Sve}, an interpretation of the transition has been attempted 
in terms of the $U(1)$ vortex dynamics of the corresponding $2D$ spin system. 
The phase transition ---  which is expected to be of i
Kosterlitz--Thouless type~\cite{KT} --- 
was demonstrated to be accompanied by restructuring of the vortex 
system~\cite{Coddington}.
The vortices are described by a
$2D$ $U(1)$ spin model 
representing the dynamics of the Polyakov line 
(see also the discussion in Ref.~\cite{Parga}). 
Approaching the transition temperature, vortices and anti--vortices start to 
form bound states. In the high temperature phase no unbound vortices and 
anti--vortices are left. 

In the present paper we  discuss an interpretation of deconfinement 
starting from the confinement picture outlined above, in terms of magnetic 
monopoles. 
The confining plasma of the monopoles and anti--monopoles turns into 
the dipole plasma at the deconfinement phase transition. The dipole plasma
is inefficient to completely screen the field created by the electric 
currents running along the pair of Polyakov lines. In this case the screening 
mass vanishes while the 
magnetic susceptibility
of the ``medium'' is smaller than unity.  
Both monopole and vortex binding mechanisms of the deconfinement phase 
transition have
been discussed for $3D$ finite temperature Georgi--Glashow model in 
Refs.\cite{AgasianZarembo} and \cite{Alex}, respectively.

In the finite temperature case,
strictly speaking, there is a problem to call all the fields
``magnetic'' as we did above when we summarized the zero temperature case.
Similar to Ref.~\cite{Coddington}, the confinement aspect itself will
be illuminated in terms of the $U(1)$ valued Polyakov lines in the third 
direction and of Polyakov line correlators representing pairs of charges 
separated in $2D$ space. In $(2+1)D$ there is no symmetry anymore between the 
three components of the field strength tensor. The closest relative of the 
true magnetic field is $F_{12}$ distinct from the others, 
while there is still a symmetry between $F_{13}$ and $F_{23}$. With this
distinction in mind one can conditionally call them the ``magnetic'' 
and ``electric'' components of the field strength tensor, respectively. 
As long as one does not introduce external fields, even at finite 
temperature there is no need to distinguish between them. The sources 
of the respective fluxes will be simply called ``monopoles'' 
or ``magnetic charges'' in the following.

The binding of the monopoles 
is not isotropic.
It happens mainly 
in the $2D$ space direction due to the 
logarithmic potential between the monopoles separated by a large spatial 
distance.
As a consequence of the
periodic boundary conditions in the temporal direction the force between 
the monopoles and anti-monopoles vanishes at half of the temporal extent. 
Therefore the potential in the temporal direction is weaker than in the spatial 
directions. As a consequence, the spatial size of the monopole bound state is 
expected to be smaller than the size in the temporal direction. This means that
the dipoles are dominantly oriented parallel or antiparallel to the 
$3^{\mathrm{rd}}$ direction.
The monopole deconfinement scenario raises the question whether the monopole 
properties
(such as pairing and orientation) could be influenced by an eventual 
external field. This
aspect  will be addressed in a companion paper. 

With or without external field, the deconfining mechanism by monopole pairing 
seems to have interesting counterparts in more realistic gauge theories.
The formation of monopole pairs is qualitatively similar to the binding of 
instantons in instanton molecules with increasing temperature in QCD suggested 
to be responsible for chiral symmetry restoration~\cite{Shuryak}.
In the electroweak theory, the formation of Nambu monopole--anti--monopole 
pairs, 
a remnant from a dense medium of disordered $Z$--vortices and Nambu monopoles 
which characterizes the high--temperature phase, is accompanying the transition 
towards the low--temperature phase~\cite{EW}. 
Note also, that a dipole vacuum, although not confining, 
still has a non--perturbative nature~\cite{QEDdipoles}.

The plan of the paper is as follows. In Section~\ref{sec:physics} we estimate 
the critical coupling of the confinement--deconfinement phase transition based 
on 
a monopole binding model for a finite lattice. In the next Section the 
transition  is numerically located for 
a lattice size $32^2\times8$ and confinement 
properties are studied. We present various monopole properties including dipole
formation based on a cluster analysis in Section~\ref{sec:monopole_properties}. 
We study the relation of the monopoles and dipoles to the phase transition in 
Section~\ref{sec:remaining}. We briefly summarize our results in the last 
Section.  

\section{Some heuristic considerations}
\label{sec:physics}

In $3D$ compact electrodynamics there are monopoles interacting via the Coulomb 
potentials,
\beqn
S = \frac{g^2_m}{2} \sum\limits_{a,b} q_a \, q_b \, V_T(\vec x_a - \vec x_b)\,,
\eeqn
where $q_a$ and $\vec x_a$ are, respectively, the charge 
(in units of the elementary monopole charge, $g_m = 2 \pi \slash g_3$, 
where $g_3$ is the three dimensional coupling constant) and the 
$3D$ position vector of the $a^{\mathrm{th}}$ monopole. 
The subscript $T$ indicates that the interaction potential $V_T$ 
eventually depends on the temperature.

At zero temperature the monopoles are randomly located in the Euclidean 
${\cal R}^3$ space and the classical interaction potential between the monopole 
and anti--monopole is inversely proportional to the distance $R$ between the 
objects, $V_0(R) = - {(4\pi \, R)}^{-1}$. 
At finite temperature $T$ the monopoles live in the 
${\cal R}^2 \times {\cal S}_1$ space (with ${\cal S}_1$ being a circle of
perimeter $T^{-1}$) and the interaction potential gets modified. 
At small separations between monopole and anti-monopole the interaction is 
zero--temperature like, 
$V_T({\mathbf{x}},z) = V_0(\sqrt{{\mathbf{x}}^2 + z^2}) + \dots$,
where $\vec x=(x,y,z)=(\mathbf{x},z)$.
At large spatial separations $\mathbf{x}$ the potential between
monopoles is essentially two--dimensional~\cite{AgasianZarembo},
\beqn
  V_T({\mathbf{x}}, z) = - 2 \, T \, \ln |{\mathbf{x}}| + \dots
  \,,\quad |{\mathbf{x}}| T \gg 1\,.
  \label{int:T}
\eeqn
However, the interaction between monopoles separated by a distance $z$
in the third (temperature) direction is of the $3D$ Coulomb type for small 
spatial
inter--monopole distances, $|{\mathbf{x}}| T \ll 1$: $V_T({\mathbf{x}}, z) = -
{(4\pi \, z)}^{-1}$, $z T \ll 1$. 
The force between monopoles and anti-monopoles at a distance $z = 1/ (2 T)$ 
vanishes due to periodicity in the temperature direction. Thus one
might expect that at 
finite temperature the monopoles form magnetically
neutral states which are bounded in the spatial directions. However, the
dynamics of the monopoles in the temperature direction is not restricted by a
logarithmic potential. 

Thus, at zero temperature the system exists in the form of a Coulomb gas of
magnetic monopoles and anti--monopoles.  In this phase the medium confines 
electric charges~\cite{Polyakov}. As temperature increases, 
the three--dimensional Coulombic potential turns into a two--dimensional 
logarithmic potential for spatial monopole interactions. The monopoles and 
anti--monopoles become weakly confined and form more and more dipole bound 
states.
The dipoles have a finite average spatial size (the distance between the 
magnetically oppositely charged constituents) which is a
decreasing function of the temperature since the interaction potential
between the particles rises as temperature increases, {\it cf.} eq.~\eq{int:T}. 

In the low temperature regime, this dipole size would be still larger than the 
average distance between the particles inside the plasma, and therefore only a 
small fraction of 
monopoles residing in 
actual dipoles is mixed with an weakly correlated 
monopole--anti--monopole component. At sufficiently large temperature, however, 
the typical dipole size becomes smaller than the interparticle distance
in the plasma and the system turns into a pure dipole plasma. The confinement 
property is closely related to the Debye mass generation effect which is absent 
in the pure dipole plasma~\cite{no-screening}. As a consequence, the 
confinement 
of electrically charged particles disappears. The system experiences a 
confinement--deconfinement phase transition due to the monopole binding 
mechanism.

One can use these heuristic arguments to estimate the phase transition 
temperature. In continuum theory this analysis was done in 
Ref.~\cite{AgasianZarembo} where compact electrodynamics was represented as a 
limit of the Georgi--Glashow model. The phase transition in this 
theory happens at a temperature $T = g^2_3 \slash (2 \pi)$. 
This result has been obtained under the condition that the average size of the 
effective magnetic dipole is not an infrared divergent quantity as 
it is the case in the confinement phase.

However, in 
lattice gauge theory the considered quantities are all finite 
and the considerations should be modified compared to the continuum case. The
difference between monopole and dipole plasmas can only be seen if the mean
distance $\bar r$ between the constituent monopoles 
becomes comparable to the dipole size $\bar d$. 
The distance $\bar r$ can be expressed 
via the density of the monopoles $\rho$, as $\bar r = \rho^{- 1 \slash 3}$. 
Thus, the phase transition
happens when the dipole size and the 
average
distance between 
monopoles become of the same order,
\beqn
   \bar d   = \xi \, \rho^{-1 \slash 3} \,,
  \label{condition}
\eeqn
where $\xi$ is a geometrical factor of order unity.
For both quantities, $\bar d$ and $\rho$, estimates can be
easily obtained on the lattice 
while the factor $\xi$ is to be defined from a simulation.

We consider the $3D$ compact $U(1)$ gauge model on the $L^2_s \times L_t$
lattice with the action written in the Villain representation:
\beqn
  \cZ = \int\limits^{+\pi}_{-\pi} \cD \theta \, 
  \sum\limits_{n(c_2) \in \Z} \,
  \exp\Bigl\{ - \frac{\beta_V}{2} {||\dd \theta + 2 \pi n||}^2 \Bigr\}\,,
  \label{Z1}
\eeqn
where $\theta$ is the compact $U(1)$ gauge field and $n$ is the integer--valued
auxiliary tensor field variable.  $\beta_V$ is the Villain coupling constant.

To relate this to the numerical simulations, we also consider the formulation 
of the compact $U(1)$ gauge theory with Wilson action:
\beqn
  S = \beta\sum\limits_P \Bigl[1 - \cos\, \theta_P \Bigr]\,.
\label{action:Wilson}
\eeqn
The Villain coupling constant $\beta_V$ is related to the Wilson coupling
$\beta$ as follows~\cite{BanksMyersonKogut}:
\beqn
  \beta_V(\beta) = {\Biggl[2 \log
  \Biggl(\frac{I_0(\beta)}{I_1(\beta)}\Biggr)\Biggr]}^{-1}\,,
  \label{betaV}
\eeqn
where $I_{0,1}$ are the standard modified Bessel functions.

The partition function~\eq{Z1} can be rewritten in the following 
``grand ca\-no\-ni\-cal''  form, {\it i.e.} represented as a sum over monopole 
charges in the (dual) lattice cubes~\cite{BanksMyersonKogut}:
\beqn
  \cZ \propto \cZ_{mon} = \sum\limits_{m(c_3) \in \Z}
  \exp\Bigl\{ - 2 \pi^2 \, \beta_V (m,\Delta^{-1} m) \Bigr\}\,.
  \label{Z2}
\eeqn	
Here $\Delta^{-1}$ is the inverse of the Laplacian operator on an asymmetric 
lattice, $m_c$ denotes the monopole charge in the cube $c_3$. 
The inverse Laplacian for lattice sizes $L_s$, $L_t$ is given as follows:
\beqn
  \Delta^{-1} (\vec x;L_s,L_t) = \frac{1}{2 L^2_s \, L_t} 
  \sum\limits_{{\vec p}^2 \neq 0}  \frac{e^{i (\vec p,\vec x)}}{3 
  - \sum^3_{i=1} \cos p_i}\,,
\eeqn
where $p_{1,2} = 0, \dots, 2 \pi (L_s - 1) \slash L_s$ and 
$p_3 = 0, \dots, 2 \pi (L_t - 1) \slash L_t$.

In order to estimate the average distance between the monopole 
and anti--monopole constituents in a dipole state ({\it i.e.}, the dipole size)
we use the ``ca\-no\-ni\-cal'' monopole--anti--monopole (dipole) partition
function which can be easily read off from eq.~\eq{Z2}:
\beqn
  \cZ^{(2)}_{dip} = 
  {\mathrm{const}} \cdot \sum\limits_{\stackrel{x}{x^2 \neq 0}}
  \exp\Bigl\{4 \pi^2 \beta_V [\Delta^{-1}(x;L_s,L_t) - 
  \Delta^{-1}(0;L_s,L_t)]\Bigr\}\,,
  \label{dipoles:pf}
\eeqn
the sum extents over all lattice separations $x$ between monopole and 
anti--monopole. The zero distance between these objects is excluded
(since this case does not correspond to a dipole state). 
The {\it r.m.s.} dipole size $\bar d$ is given by
\beqn
  {\bar d}^2(\beta_V;L_s,L_t) = \frac{1}{\cZ^{(2)}_{dip}} \, \sum\limits_x
  x^2 \, \exp\Bigl\{4 \pi^2 \beta_V \, \Bigl[\Delta^{-1}(x;L_s,L_t)
  - \Delta^{-1}(0;L_s,L_t) \Bigr]\Bigr\}\,,
  \label{bar:d}
\eeqn
where the actual distance squared, $x^2$, is evaluated taking into account the
periodic boundary conditions of the lattice. The sums cannot be taken 
analytically.

The monopole density $\rho$ can be 
read off from eq.\eq{Z2},
\beqn
  \rho(\beta;L_s,L_t) = 2 \exp\Bigl\{ - 2 \pi^2 \beta_V(\beta) \, 
  \Delta^{-1}(0;L_s,L_t) \Bigr\}\,,
  \label{rho:theory}
\eeqn
where the dependence on the lattice geometry is indicated explicitly.
Note that in this formula 
no interaction between monopoles 
is taken into account and 
we refer to it as to ``bindingless''.
Only the local ``self--interaction'' of monopoles is accounted for 
via the Coulomb propagator $\Delta^{-1}(0)$ 
in the fugacity.
We are discussing the binding effects on the monopole density in Section~\ref{sec:remaining}.

The geometrical factor $\xi$ is to be defined from the
numerical data. To this end we assume that this factor is a constant i
quantity which does not
depend on the lattice extensions. Indeed, it gives an estimate how large 
the intra--dipole distances should be compared to the monopole density
in order 
to have the dipole field screened.
This is a quite strong assumption which, however,
turns out to be 
reasonable, as will we see below. To define the factor $\xi$ we substitute
eqs.~(\ref{bar:d}) and (\ref{rho:theory}) into eq.~(\ref{condition}) and use 
numerical values
for $\beta_c$ presented in Ref.~\cite{Coddington}. For the 
lattices $16^2\times L_t$, $L_t =
4,6,8$ we get, respectively: $\xi=0.723(58),~0.622(47)$ and $0.646(116)$. 
These numbers
coincide with each other within numerical errors. 
Taking the average over $L_t$ we get  $\bar \xi \approx 2/3$. 
In what follows we take 
\beqn
\xi = 2 \slash 3\,,
\label{xi}
\eeqn
and then solve eqs.(\ref{condition},\ref{rho:theory},\ref{bar:d}) 
with respect to the Villain
coupling $\beta_V$. Then we  finally estimate the critical Wilson coupling
$\beta^{\mathrm{th}}_c$  with the help of eq.~\eq{betaV}. 
The results for lattices of various
sizes are represented in Table~\ref{tab:beta:th}
%%%%%%%%%%%%%%%%%%%%%%%%%%%%%%%%%%%%%%%%%%%%%%%%%
\begin{table}[!htb]
\vspace{3mm}
\begin{center}
\begin{tabular}{||c|c|c|c|c|c||}
\hline
\hline
&\multicolumn{2}{c|}{$L_s=16$} & \multicolumn{2}{c|}{$L_s=32$} & $L_s=64$\\
\hline
$L_t$ & $\beta^{th}_c$  & $\beta_c$ & $\beta^{th}_c$  & 
$\beta_c$ & $\beta^{th}_c$ \\
\hline
4 & 1.87 & 1.83(2) & 2.01 &   -     &  2.10\\
6 & 2.04 & 2.08(2) & 2.26 & 2.18(3) &  2.44\\
8 & 2.12 & 2.14(5) & 2.39 & 2.30(2) &  2.62\\
\hline
\hline
\end{tabular}
\end{center}
\caption{The critical coupling constant $\beta^{th}_c$ calculated 
using eqs.~(\ref{betaV},\ref{condition},\ref{bar:d},\ref{rho:theory},\ref{xi}) 
for different lattices $L^2_s \times L_t$ compared to lattice Monte Carlo 
results of Ref.~\cite{Coddington}. 
Note that our results for the lattice $32^2\times 8$ are slightly
higher than that of
Ref.~\cite{Coddington}, see forthcoming Sections.}
\label{tab:beta:th}
\end{table}
%%%%%%%%%%%%%%%%%%%%%%%%%%%%%%%%%%%%%%%%%%%%%%%%%%%%
and compared with pseudocritical  couplings  $\beta_c$ obtained in 
lattice simulations of
Ref.~\cite{Coddington}. The agreement between the data  
and our estimates is within $4\%$.
Thus the simple heuristic arguments based on the monopole--dipole picture work 
surprisingly well.

\section{Phase transition and confinement}
\label{sec:simulations}

We have performed our numerical study of $(2+1)D$ compact electrodynamics using 
the Wilson action (\ref{action:Wilson}). The lattice coupling $\beta$ is related
to the lattice  spacing $a$ and the continuum coupling constant
$g_3$ of the $3D$ theory as follows: 
\beqn
  \beta = \frac{1}{a\, g^2_3}\,,
  \label{beta}
\eeqn
Note that in three dimensional gauge theory the coupling constant $g_3$ has 
dimension ${\mathrm{mass}}^{1 \slash 2}$.

The lattice corresponding to the finite temperature is asymmetric, $L^2_s\times
L_t$, $L_t < L_s$. In the limit $L_s \to \infty$ the ``temporal'' extension 
of the
lattice $L_t$ is related to the physical temperature, $L_t = 1 \slash (T a)$.
Using eq.~\eq{beta} the temperature is given via the lattice parameters as 
follows:
\beqn
  \frac{T}{g^2_3} = \frac{\beta}{L_t}\,.
  \label{temp}
\eeqn
Thus, at fixed lattice size lower (higher) values of the lattice coupling 
constant $\beta$ correspond to lower (higher) temperatures.

Our simulations have been performed mainly on a $32^2 \times 8$ lattice. 
We do not intend to study in the present paper finite size scaling 
aspects of this model.
The local 
Monte Carlo algorithm is based on a 5--hit Metropolis update 
sweep followed by a  microcanonical sweep. For better ergodicity, in particular 
in the presence of an external field (considered in a companion paper), also 
global updates are included.  Following the ideas of 
Ref.~\cite{DamgaardHeller}, 
the global refreshment step consists in an attempt to add an additional 
unit of flux with randomly chosen sign in a direction randomly selected
among the three, to the dynamical gauge field subject to a global Metropolis 
acceptance check.  

For example, one unit of flux in $ij$ plane is introduced with the help of the
following gauge field shift~\cite{DamgaardHeller} 
$\theta_i \rightarrow {[\theta_i + \tilde{\theta}_i]}_{{\mathrm{mod}} 2\pi} $:
\beqn
  \tilde{\theta}_j & = & \frac{\pi}{L_i} (2 x_i - L_i - 1)\,, \quad 
  \tilde{\theta}_j = 0\,\,\,\, {\mathrm{for}} \,\,\,\, x_j \neq L_j\,, 
  \nonumber\\
  \tilde{\theta}_i & = & \frac{2 \pi}{L_i \, L_j} \, (1-x_j)\,,\quad
  \tilde{\theta}_k = 0\,,\,\,\, k \neq i,j\,. \nonumber
\eeqn
The acceptance rate of the global step changes within the considered $\beta$ 
range from roughly 0.7 (confinement phase) to 0.2 (deconfinement phase).  
One total Monte Carlo update cycle  consists of two combined local Metropolis 
and microcanonical sweeps (requiring an acceptance rate of 0.5 for the 
Metropolis step) and the global update described above.

In order to localize the deconfinement transition, it is convenient to study the
expectation values of the two bulk operators,
\beqn
  \la |L| \ra = \frac{1}{L^2_s} \, \la |\sum\limits_{\mathbf x} L(\mathbf x)|
  \ra\,, \quad 
  \la |L|^2 \ra = \frac{1}{L^2_s} \, \la |\sum\limits_{\mathbf x}
  L(\mathbf x)|^2 \ra  \,,
  \label{zero:bulk:pol:loop}
\eeqn
constructed from the Polyakov loop,
\beqn
  L(\mathbf x) = \exp \Bigl\{i \sum\limits_{z=1}^{L_t} \theta_3(\mathbf x,z)
  \Bigr\}\,, 
\eeqn
here $\mathbf x = (x, y)$ is a two--dimensional vector.
In the deconfinement phase the quantity $|L|$ is of the order of unity, while 
in the confinement phase it is close to zero in a finite volume and vanishes 
in the infinite volume limit.
 
\begin{figure*}[!htb]
\begin{center}
\begin{tabular}{cc}
\epsfxsize=7.0cm \epsffile{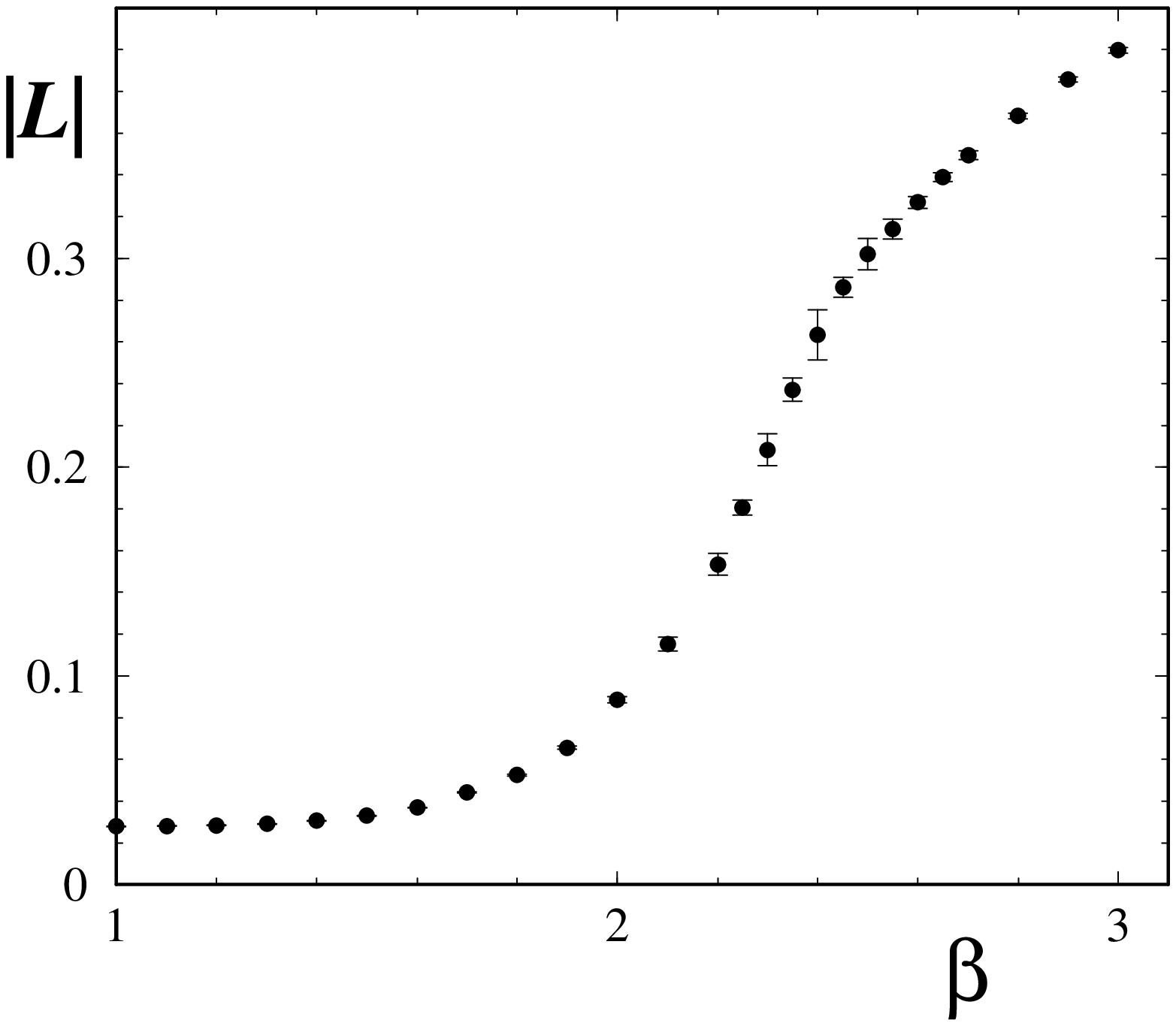}  &  
\epsfxsize=7.0cm \epsffile{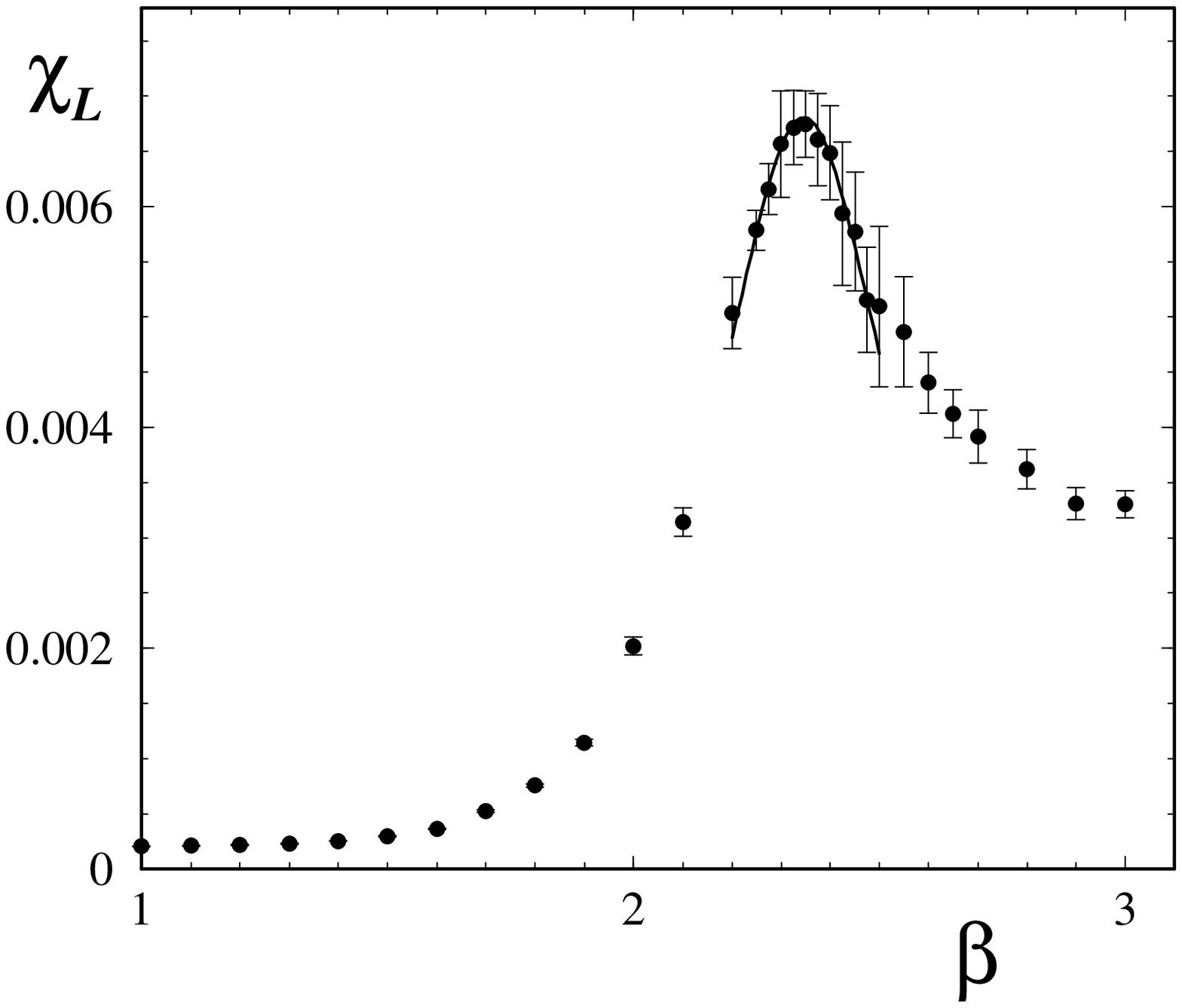} \\
(a) & \hspace{1.5cm}  (b) \\
\end{tabular}
\end{center}
\vspace{-0.5cm}
\caption{(a) The expectation value of the absolute value of the average 
Polyakov loop~\eq{zero:bulk:pol:loop} and (b) its 
susceptibility~\eq{zero:pol:sus} {\it vs.} $\beta$.}
\label{fig:zero:bulk:polyakov}
\end{figure*}

The behaviour of the expectation value of the Polyakov loop {\it vs.} lattice 
coupling $\beta$ is shown in Figure~\ref{fig:zero:bulk:polyakov}(a). 
%%%%%%%%%%%%
%%%%%%%%%
The low temperature phase, $\beta < \beta_c$ corresponds to the confinement 
phase, while the high temperature phase is deconfining.

The susceptibility of the Polyakov loop
\beqn
  \chi_L = \la {|L|}^2 \ra - {\la |L| \ra}^2 
  \label{zero:pol:sus}
\eeqn
is shown in Figure~\ref{fig:zero:bulk:polyakov}(b). The peak of the Polyakov
loop susceptibility corresponds to the pseudocritical $\beta_c$ of the 
deconfinement phase transition. We have fitted the susceptibility near its 
maximum by the following function:
\beqn
  \chi^{\mathrm{fit}}_L(\beta) = \frac{c^2_1}{c^2_2 + (\beta - \beta_c)^2}\,,
\eeqn
where the critical coupling was estimated to be $\beta_c = 2.346(2)$ which is
quite close to the result of Ref.~\cite{Coddington}. The best fit is shown in
Figure~\ref{fig:zero:bulk:polyakov}(b) by a solid line.

\begin{figure*}[!htb]
\begin{center}
\begin{tabular}{cc}
\epsfxsize=7.0cm \epsffile{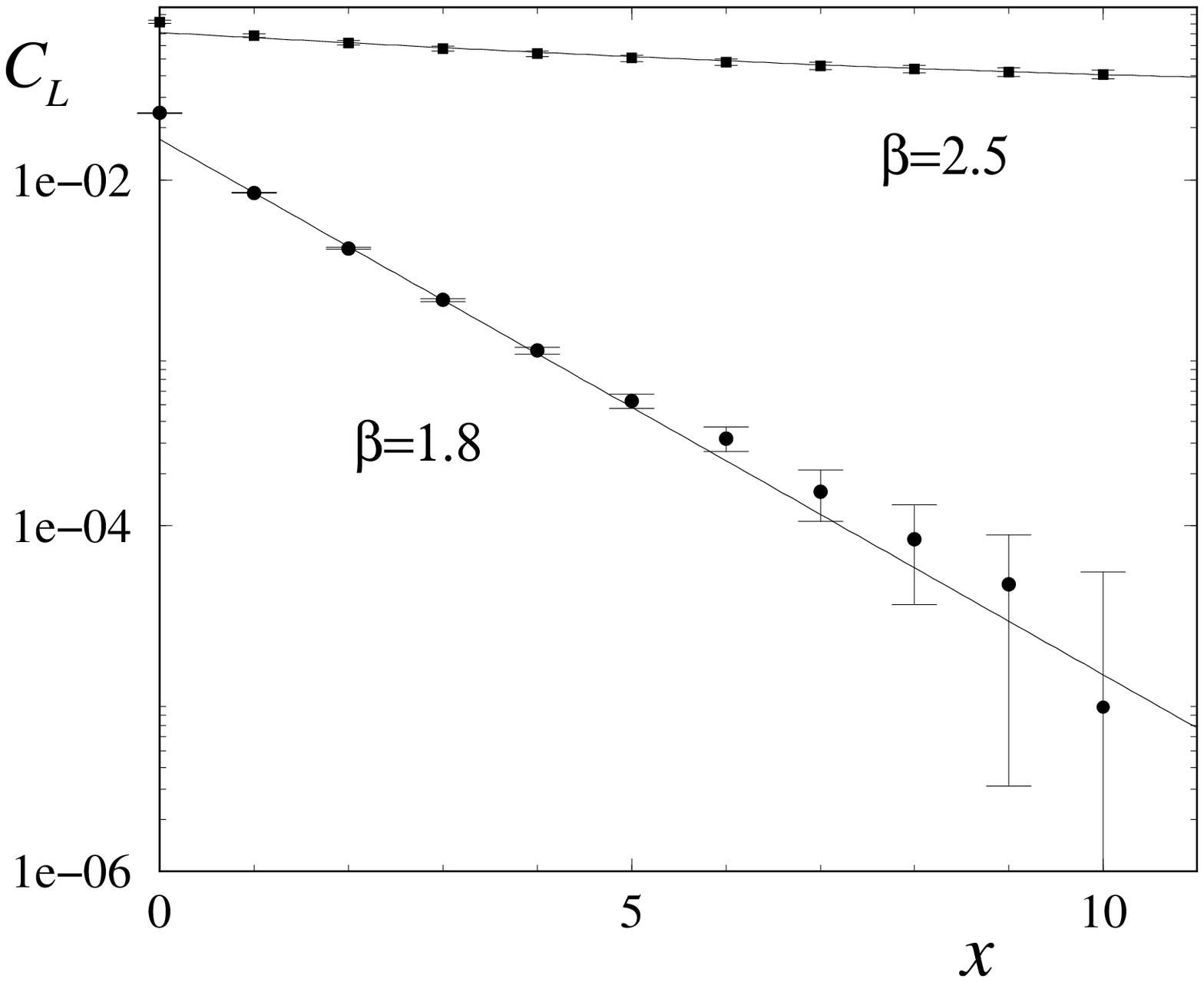}  &  
\epsfxsize=7.0cm \epsffile{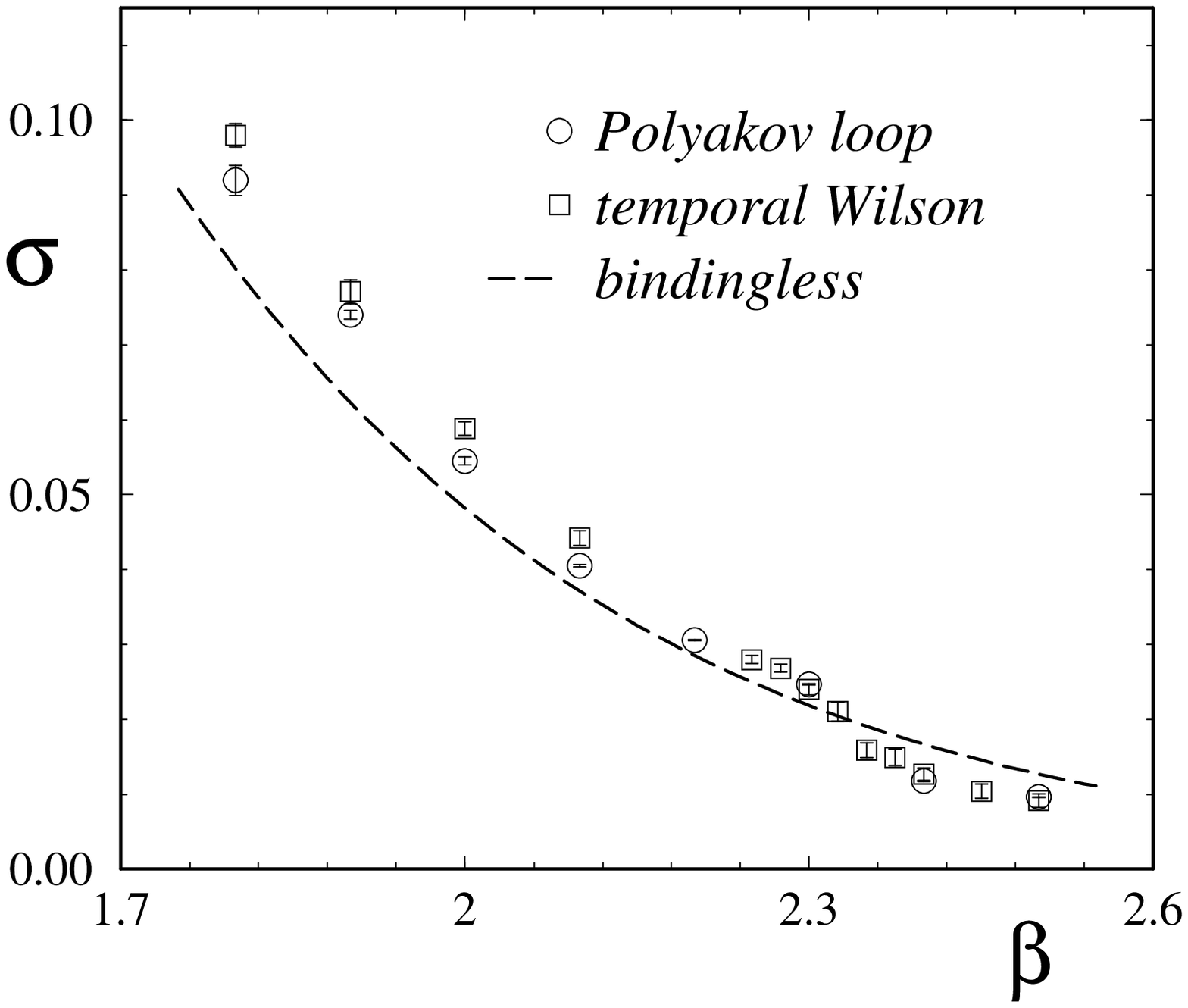} \\
(a) & \hspace{1.5cm}  (b) \\
\end{tabular}
\end{center}
\vspace{-0.5cm}
\caption{(a) Fit of the Polyakov plane--plane correlator~\eq{pol:pol:corr}
using eq.~\eq{fit:fun} in the confinement, $\beta=1.8$, and deconfinement, 
$\beta=2.5$, phases. (b) String tensions as functions of $\beta$ compared with 
the  bindingless  theoretical result~\eq{zero:temp:string}.}
\label{fig:string:tension}
\end{figure*}

To calculate the string tension we use ``plane--plane'' correlators of two 
Polyakov loops. In addition, averages of temporal Wilson loops have been 
studied, too.  
More precisely, in $(2+1)D$, we define first sums of the Polyakov loops 
along a line parallel to a spatial lattice axis ({\it e.g.} in
the $y$ direction):
\beqn
  L_{\mathrm{plane}}(x) = \sum\limits_{y=1}^{L_s} L(x,y) \, . 
\eeqn
The correlator of the plane--plane correlators may be written as a sum of
point--point correlation functions,
\beqn
  \la L_{\mathrm{plane}}(0) L^{*}_{\mathrm{plane}}(x)\ra 
  = \sum\limits_{y_{1,2}=1}^{L_s} \la L(0,y_1) \, L^{+}(x,y_2) \ra\,.
  \label{pol:pol:corr}
\eeqn
The form of this correlator is expected to be:
\beqn
  \la L_{\mathrm{plane}}(0) L^{*}_{\mathrm{plane}}(x)\ra = {\mathrm{const}} 
  \cdot \cosh\Bigl[\sigma L_t \, \Bigl( x - \frac{L_s}{2} \Bigr)\Bigr]\,,
  \label{fit:fun}
\eeqn
where $\sigma$ is the ``temporal'' string tension. 
In Figure~\ref{fig:string:tension}(a)
%%%%%%%%%
%%%%%%%%%%
we show the fit of the Polyakov plane--plane correlator~\eq{pol:pol:corr}
by this fitting function in the confinement ($\beta=1.8$) 
and deconfinement ($\beta=2.5$) phases, respectively.

In Figure~\ref{fig:string:tension}(b) we present the fitted string 
tensions as  function of $\beta$. Above $\beta_c$ the string tension quickly 
drops down but stays non-zero due to finite volume effects.
The temporal string tensions obtained using either
the Polyakov loop plane--plane correlators or the temporal Wilson loop
averages roughly coincide with each other.
The dashed curve represents the theoretical prediction
for the string tension~\cite{BanksMyersonKogut,Ambjorn}:
\beqn
  \sigma(\beta) = \frac{4 \sqrt{2}}{\pi \sqrt{\beta_V(\beta)}} \exp\Bigl\{
  - \pi^2 \, \beta_V(\beta) \, \Delta^{-1}(0; L_s, L_t) \Bigr\}\,.
  \label{zero:temp:string}
\eeqn

Agreement between the prediction and the numerical results is 
reached only in the vicinity of the phase transition point, 
$\beta \approx 2.3$. 
In order to understand these differences, we turn now to a closer investigation 
of the monopole properties.

\section{Properties of the monopole--anti--monopole system}
\label{sec:monopole_properties}

\begin{figure*}[!htb]
\begin{center}
\epsfxsize=8.0cm \epsffile{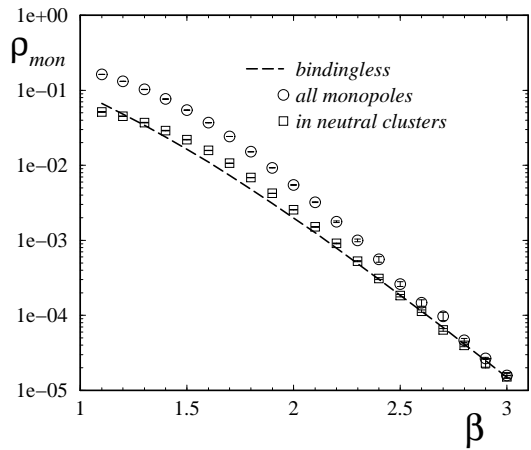}
\caption{The density of all monopoles and of 
monopoles in neutral clusters {\it vs.}~$\beta$.} 
\label{fig:mon}
\end{center}
\end{figure*}

\begin{figure*}[!htb]
\begin{center}
\begin{tabular}{cc}
\epsfxsize=7.5cm \epsffile{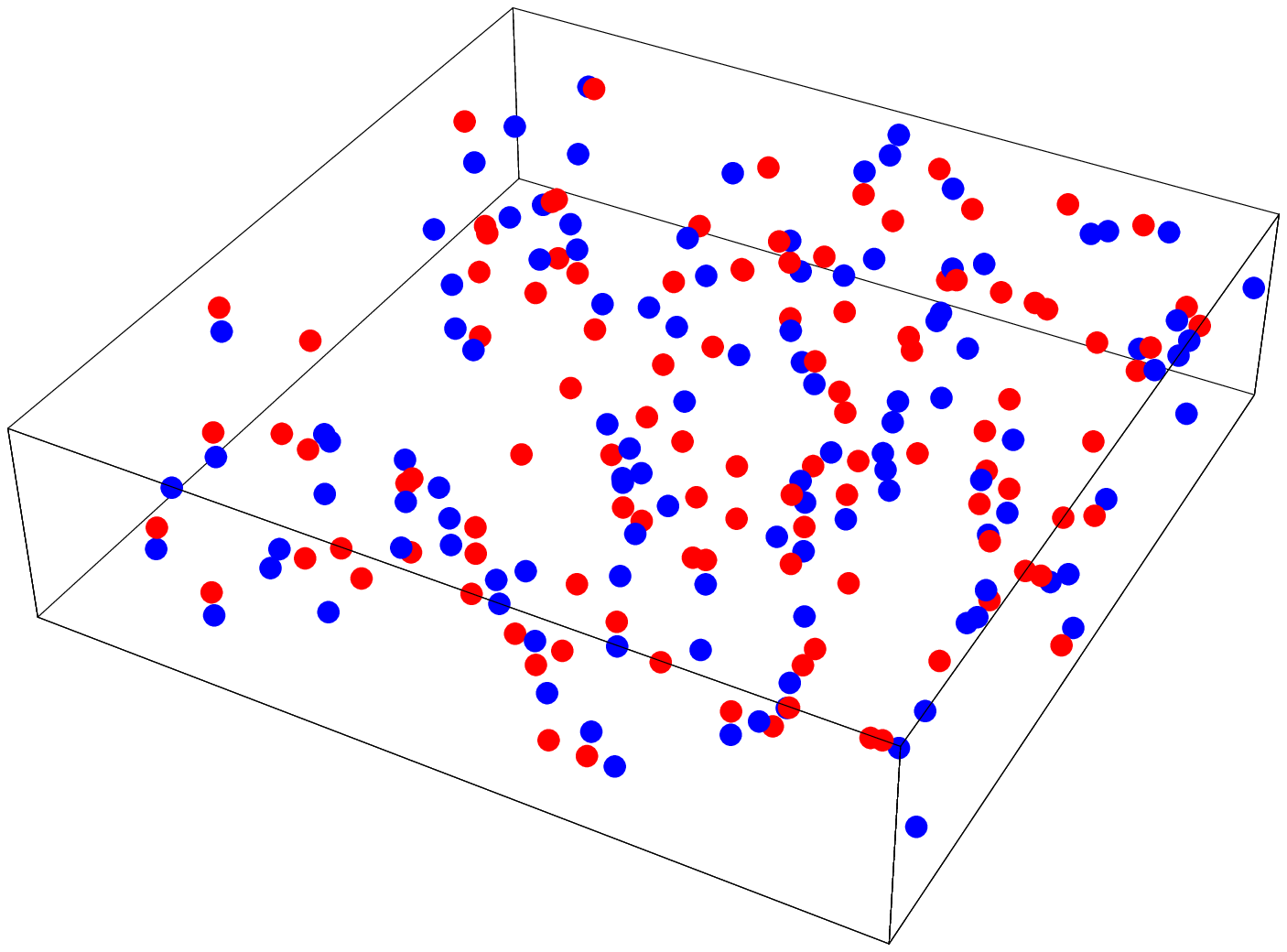} &
\epsfxsize=7.5cm \epsffile{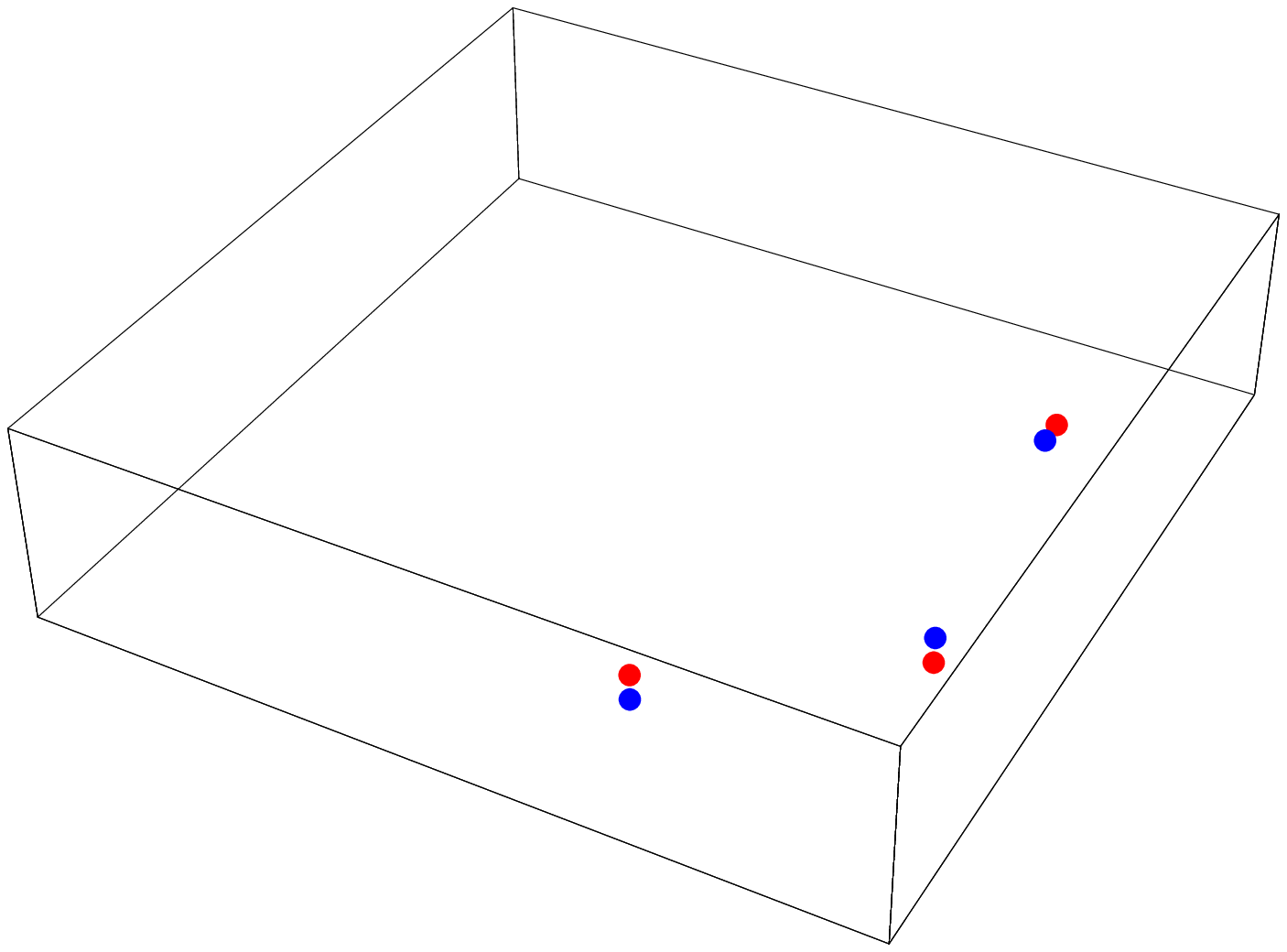} \\
(a) & \hspace{1.5cm}  (b) \\
\end{tabular}
\end{center}
\vspace{-0.5cm}
\caption{Typical monopole configurations for (a) the confinement phase ($\beta
= 1.6$) and (b) the deconfinement phase ($\beta=2.5$) \,.}
\label{fig:configurations}
\end{figure*}

The basic quantity describing the behaviour of the monopoles is the
monopole density, $\rho = \sum_c |m_c|/(L_s^2 L_t)$, where $m_c$ is the integer
valued monopole charge inside the cube $c$ defined in the standard
way~\cite{DGT}:
\beqn
  m_c = \frac{1}{2\pi} \sum\limits_{P \in \partial c} {(-1)}^P \, {[
  \theta_P]}_{\mathrm{mod} \, 2 \pi}\,,
\eeqn
where the plaquette orientations 
relative to the boundary of the cube 
are taken into account.
The density of the total number of monopoles is a decreasing function of the
lattice coupling $\beta$ (or the temperature) as it is shown in 
Figure~\ref{fig:mon}
%%%%%%%%%%%%%%%%%%%%%%%%%%%%%%%%%%%
by circles.
At high temperatures (large $\beta$) the monopoles are dilute and form 
dipole bound states. 
Typical monopole configurations in both phases are visualized in 
Figure~\ref{fig:configurations}.
%%%%%%%%%%%%%%%%%%%%%%%%%%%%%

In Figure~\ref{fig:mon} we show 
by the dashed line the density of the 
monopoles calculated using eq.~\eq{rho:theory}
for comparison. 
As in the case of the string tension, the predicted monopole density
is in agreement with the numerical data 
only near $\beta_c$.

Equation~\eq{rho:theory} is based on the single monopole contribution to
the partition function, thus it does not take into account pairing of the
monopoles. The effect of the constituent monopole pairing (dipole
formation) due to finite temperature can explain the deviations from the
bindingless case seen in this Figure. In the confinement phase the density
of the monopoles is larger than the prediction of eq.~\eq{rho:theory}.
Indeed, we expect that the formation of the bound state decreases the
total energy (action) of the chosen monopole and anti--monopole. As a
result binding favours the creation of additional monopoles by quantum
fluctuations. This tendency increases with larger $\beta$, however the
cost of creating new monopoles grows, too.

Note that the entropy of the bound state is smaller than the
entropy of a free monopole and an anti--monopole. However the entropy
effect does not seem to change essentially near the phase transition.

We remind the reader that on the classical level the dipoles are formed both in 
the confinement phase and in the deconfinement phase due to logarithmic
potential between the monopoles. However, 
at low temperatures the dipole size is larger than the average distance 
between the
monopoles and, therefore, the dipole formation does not destroy confinement.

One can analyse the monopole pairing  studying the cluster structure of 
the monopole ensemble extracted from the Monte--Carlo configurations. For our
purposes, clusters are defined as follows: clusters are connected groups of
monopoles and anti--monopoles, where each object
is separated from at least one neighbour belonging to the same cluster by
a distance less or equal than $R_{\mathrm{max}}$. In the following we use
$R^2_{\mathrm{max}}=3~a^2$ which means that neighbouring monopole 
cubes should share at least one single corner\footnote{
In Ref.\cite{TeperSchram} a similar definition has been used to investigate 
tightly packed 
clusters with $R_{max}=a$. In our case the condition for the cluster is more
relaxed.}.
Note that the increase of the coupling constant
leads not only to an increase of the temperature, eq.~\eq{temp}, but to a 
decrease of
the lattice spacing $a$ as well, eq.~\eq{beta}. Thus at different $\beta$ the
same characteristic distance $R_{\mathrm{max}}$ corresponds to different 
physical scales. 
Therefore our results below are of only qualitative nature.

A monopole cluster is neutral if the charges of the constituent monopoles sum 
up to zero. We show the density of monopoles belonging to neutral clusters as a
function of $\beta$ in Figure~\ref{fig:mon} by squares. The difference between 
this density and the total monopole density amounts to a factor three at
$\beta \approx 1$ and becomes smaller at larger $\beta$. At large $\beta$
(entering the deconfinement phase) approximately every second monopole or
anti--monopole belongs to a neutral cluster.  At still larger $\beta$'s almost
all monopoles are in neutral clusters. 

\begin{figure*}[!htb]
\begin{center}
\begin{tabular}{cc}
\epsfxsize=7.0cm \epsffile{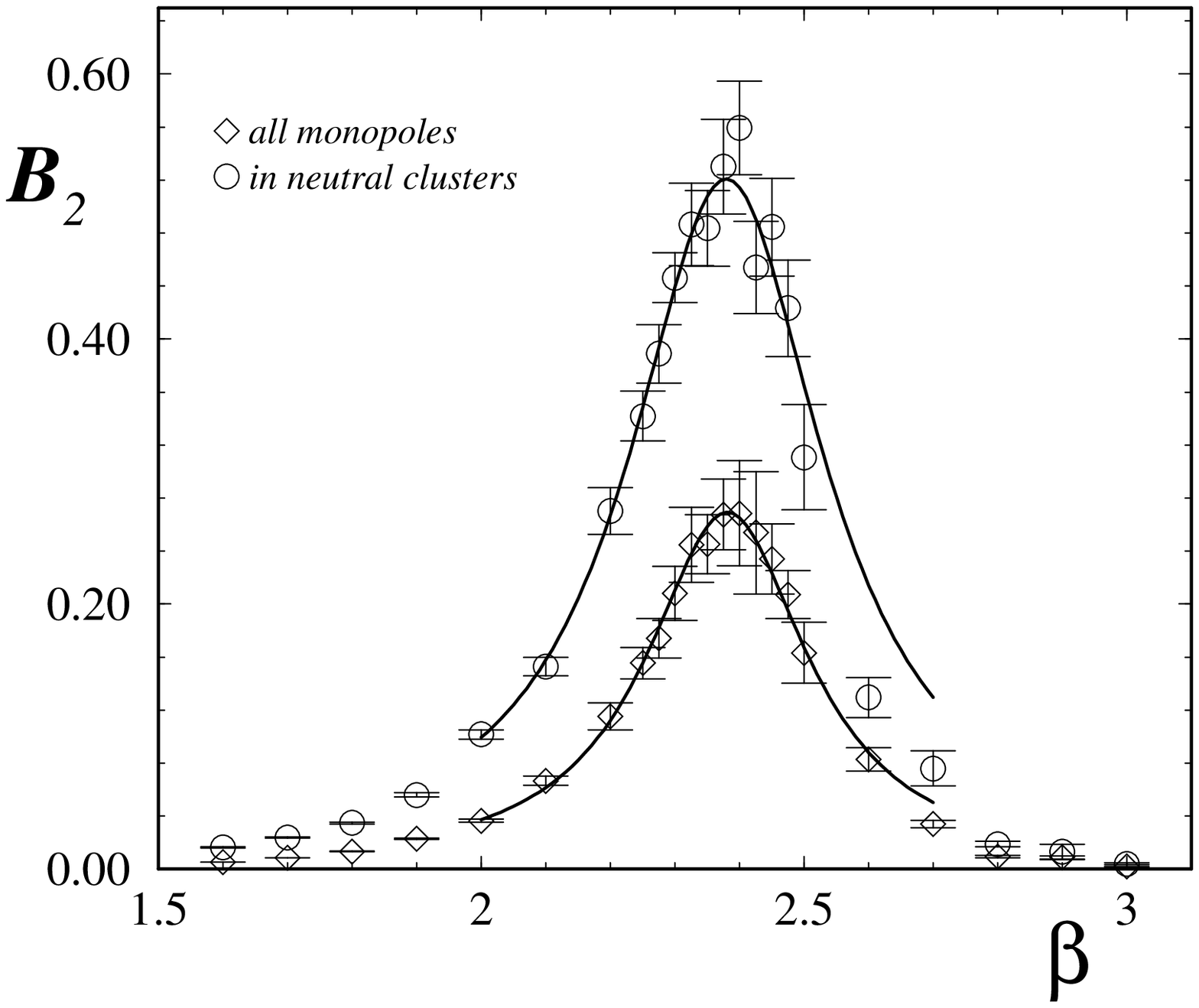} & 
\epsfxsize=7.0cm \epsffile{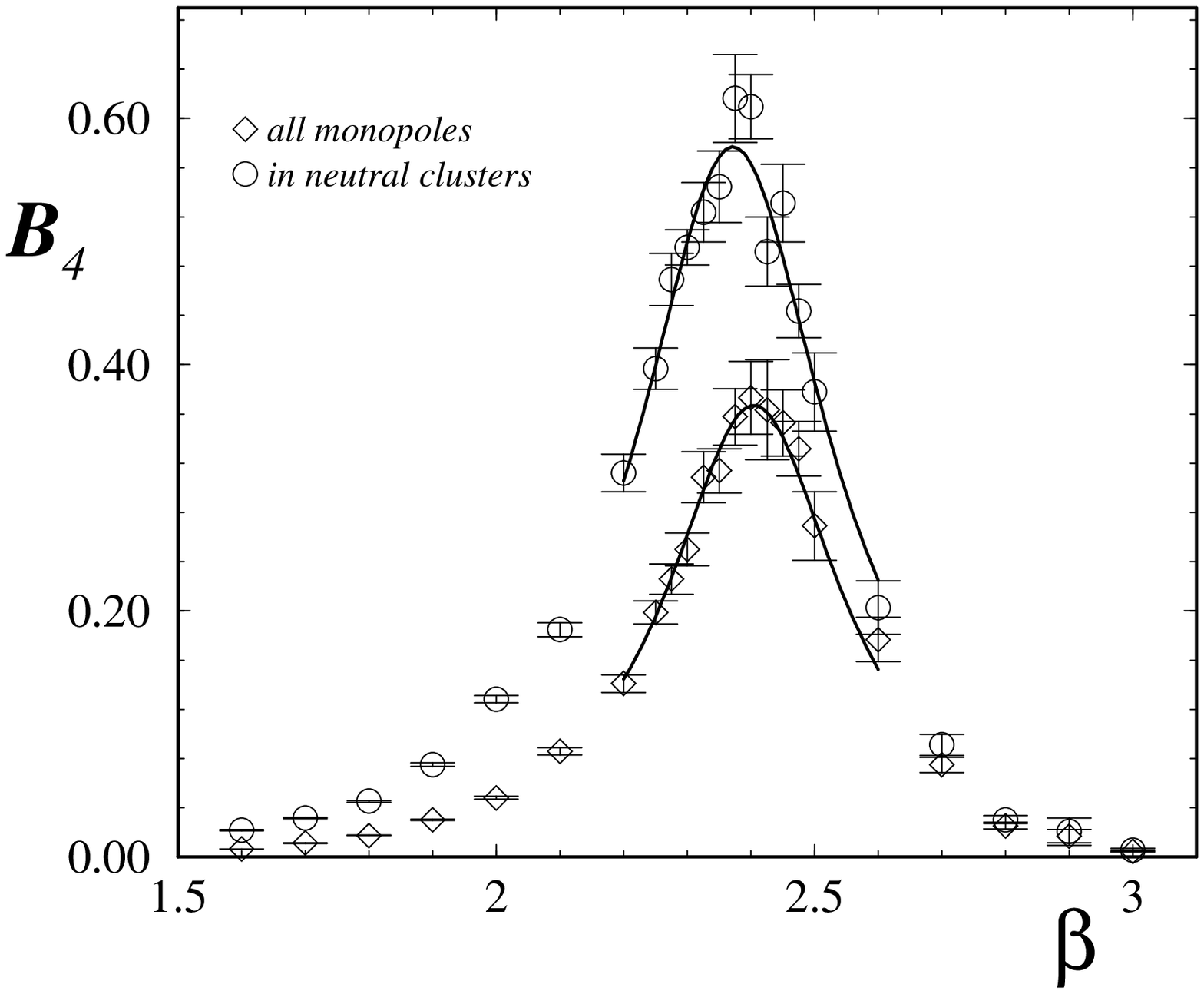} \\
(a) & \hspace{1.5cm}  (b) \\
\end{tabular}
\end{center}
\vspace{-0.5cm}
\caption{The second (a) and the fourth Binder (b) cumulants according 
to~\eq{Binder1} for the total (anti--)monopole density and the corresponding 
densities enclosed in neutral clusters. The fits are shown as solid lines.}
\label{fig:b2:b4:all}
\end{figure*}

We are confident that the fluctuation of monopole numbers signal the 
deconfining phase transition. 
This is demonstrated studying the second and (modified) 
fourth Binder cumulants of the total  number of monopoles and anti--monopoles 
[and of the number of (anti--)monopoles being  part of neutral clusters]. 
We present in Figure~\ref{fig:b2:b4:all}
%%%%%%%%%%%%%%%%%%%%%%%%%%%%
the cumulants, with $M$ denoting the respective number :
\beqn
  B_2 & = & \frac{\la M^2 \ra}{{\la M \ra}^2} - 1 \\
  B_4 & = & \frac{\la M^4 \ra}{3 {\la M^2 \ra}^2} - \frac{1}{3} \,,
  \label{Binder1}
\eeqn
Similarly to the Polyakov line susceptibility these quantities are suitable 
to localize the deconfining phase transition. We fit these cumulants by 
\beqn
  B_n^{\mathrm{fit}}(\beta) = 
  \frac{c^2_1}{c^2_2 + (\beta - \beta_c)^2}\,, \quad n =2,4\,.
  \label{Binder2}
\eeqn
The fits are shown by solid lines in Figure~\ref{fig:b2:b4:all} and the 
results for the pseudocritical couplings  $\beta_c$ are given in 
Table~\ref{tab:beta:exp}.
\begin{table}[!htb]
\begin{center}
\begin{tabular}{|lll|}
\hline
cumulant 	&    2nd  	&   4th  \\
\hline
total		&    2.380(3)   &   2.404(4) \\
neutral		&    2.379(5)   &   2.372(3) \\
\hline
\end{tabular}
\end{center}
\caption{Pseudocritical couplings  $\beta_c$ from 
the fits to the Binder cumulants~(\ref{Binder1},\ref{Binder2}).} 
\label{tab:beta:exp}
\end{table}

\begin{figure*}[!htb]
\begin{center}
\begin{tabular}{cc}
\epsfxsize=7.0cm \epsffile{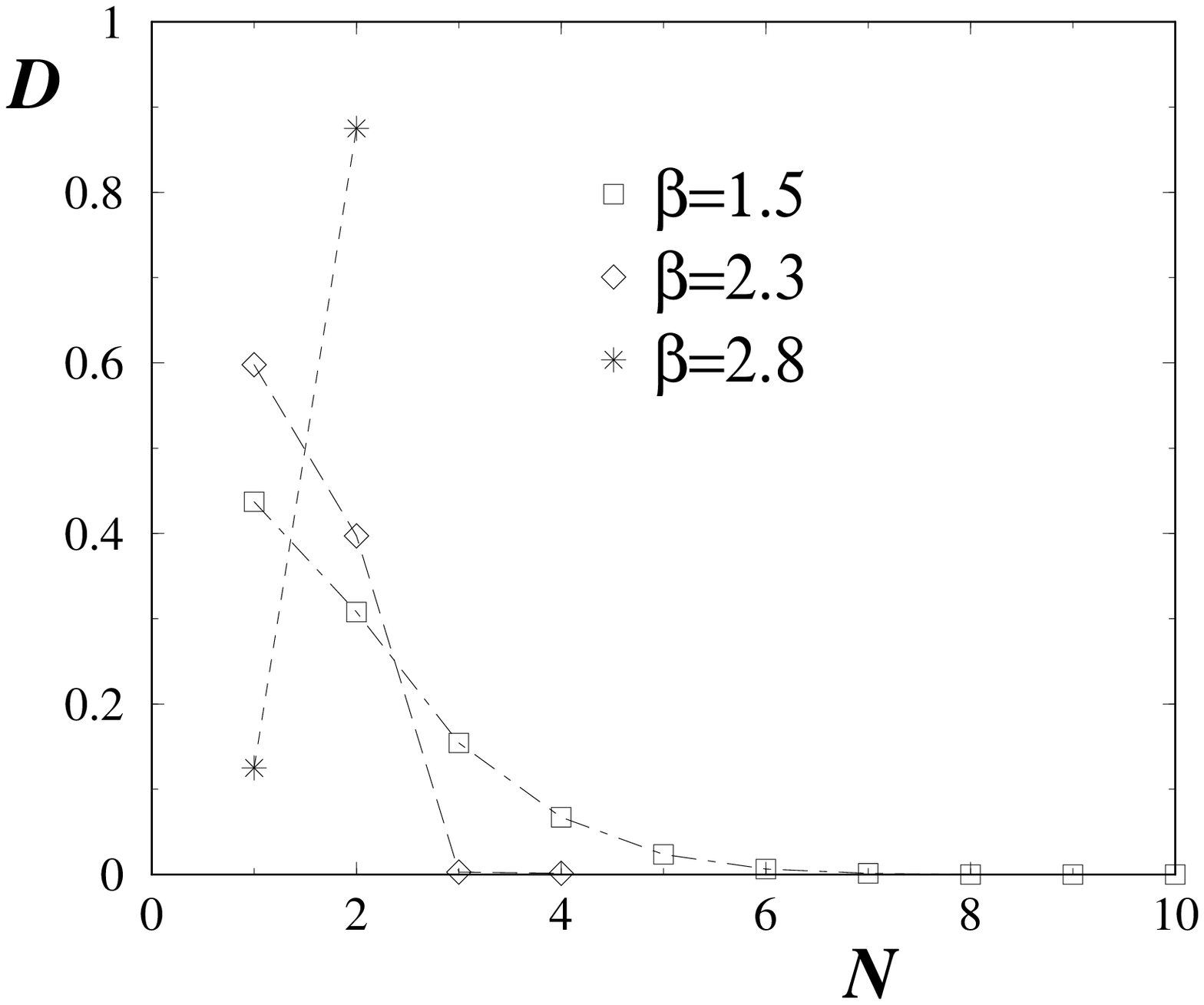} & 
\epsfxsize=7.0cm \epsffile{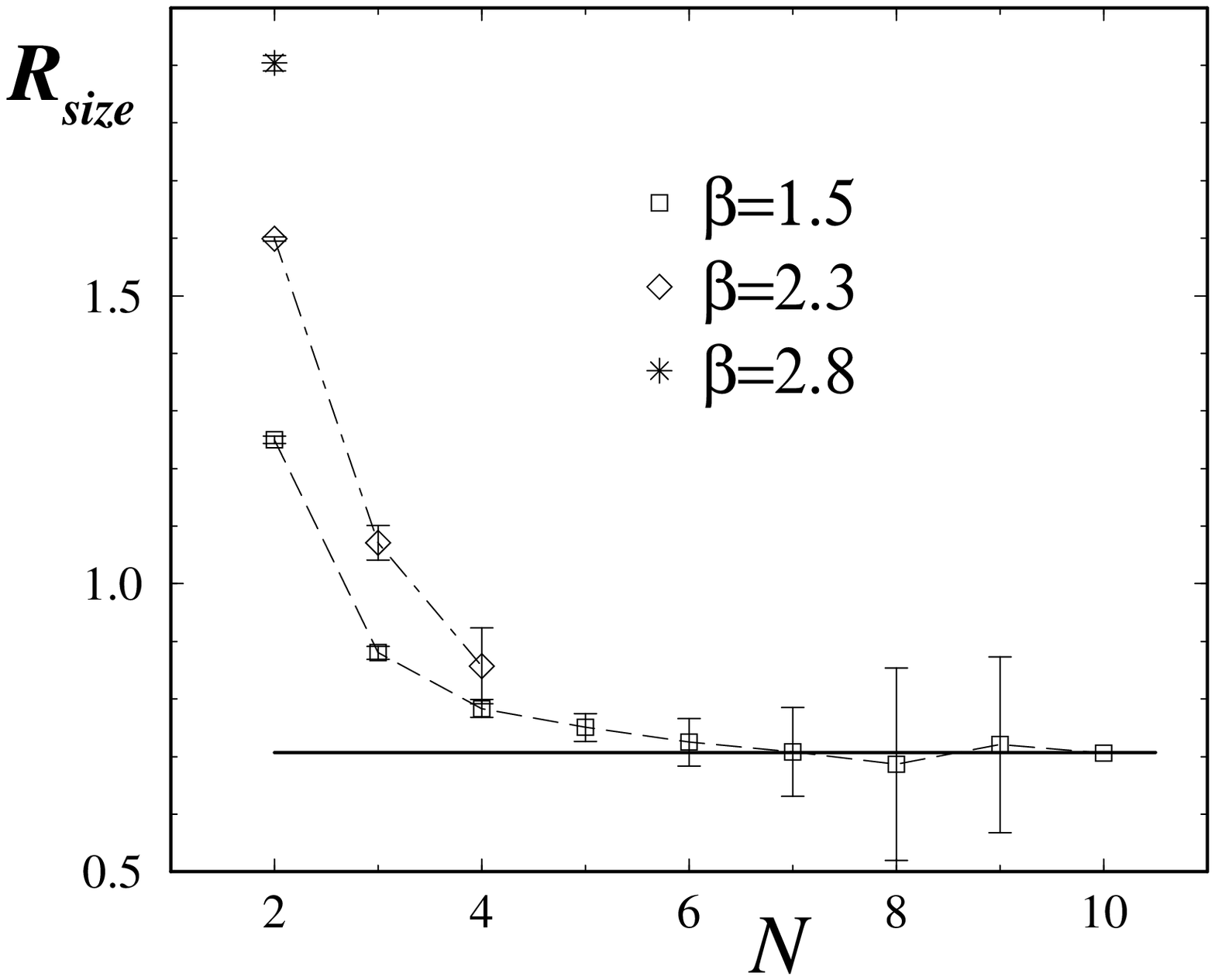} \\
(a) & \hspace{1.5cm}  (b) \\
\end{tabular}
\end{center}
\vspace{-0.5cm}
\caption{(a) The cluster structure at various of the coupling constant
$\beta$. Cluster distribution is shows as a function of  the number of
constituent monopoles inside clusters, $N$. (b) The cluster shape function, 
eq.~\eq{along}, for various $\beta$.} 
\label{fig:clusters}
\end{figure*}

Some details on the cluster structure at various values of $\beta$ can be 
seen in Figure~\ref{fig:clusters}(a). 
%%%%%%%%%%%%%%%%%%%%%%%%%%%%%
We show the fraction of monopoles and anti--monopoles being part of clusters
of size $N$.  The cluster size is the number of monopoles and anti-monopoles 
which belong to the given cluster. There is no separation according to the 
cluster's charge.  In the confinement phase, $\beta=1.5$, the fraction of 
monopoles is slowly decreasing with the cluster size $N$. The percentage of
isolated (anti--)monopoles ($N=1$ clusters) amounts to roughly 45 \% while 
clusters (with a size up to $N=10$) contain the rest.  

At the phase transition point ($\beta\approx 2.3$) the number of 
(anti--)monopoles enclosed in larger clusters drops drastically. The monopole 
vacuum is composed mostly of individual (anti--)monopoles (60 \%) and dipoles 
(40 \%). This observation can be reconciled with our theoretical expectation 
that all monopoles must become paired only if we accept that the
``unpaired'' monopoles are actually part of dipoles of size bigger 
than $R_{\mathrm{max}}$. 
Deeper in the deconfined phase, however, at $\beta=2.8$ practically 90 \% of the
(anti--)monopoles form tight bound states with sizes smaller than
$R_{\mathrm{max}}=\sqrt{3}~a$. 

As we have discussed above, we expect that the force in the spatial directions 
is  larger than the force along the temporal direction $z$.
This fact can be qualitatively analysed with the help of the following ``cluster
sphericity'':
\beqn
  R_{\mathrm{size}}(N) = \frac{{\la |\Delta z| \ra}_N}{\sqrt{{\la
  |\Delta x| \ra}^2_N + {\la |\Delta y|\ra}^2_N }}\,,
  \label{along}
\eeqn
where ${\la | \Delta x | \ra}_N$ is the average distance from the center of 
the cluster in $x$--direction {\it etc.}
for cluster size $N$. 
If the clusters are elongated predominantly in
the temporal direction this quantity would be larger than unity, and 
smaller otherwise. We show the dependence of the sphericity $R_{\mathrm{size}}$ 
on the cluster size $N$ for various $\beta$ values in 
Figure~\ref{fig:clusters}(b). Small clusters are directed predominantly along 
the temporal direction, as expected, at all $\beta$. 
With larger $\beta$ the elongation becomes stronger. 
For large clusters the 
direction of the cluster is random, since in this case the cluster shape 
function is very close to $1 \slash \sqrt{2}$ (this directly follows from the 
definition~\eq{along}). This random limit is marked by the solid line in 
Figure~\ref{fig:clusters}. 

\section{Confinement and monopoles}
\label{sec:remaining}

We have observed that agreement between  predictions from a theory without
monopole binding and the finite temperature simulation results is reached only 
in the vicinity of the phase transition point, $\beta \approx 2.3$. 
In the confinement phase both measured temporal string tension 
and monopole density are larger compared to the 
bindingless predictions,
see Figures~\ref{fig:string:tension}(b),~\ref{fig:mon}. 

As we have discussed, it is due to monopole binding that the density of the 
monopoles is increased compared to the non-interacting case. 
However, the size of the dipoles in the confinement phase is
larger than the average distance between the ordinary monopoles 
calculated from their total density. Therefore, the monopoles bound in dipoles 
due to classical logarithmic potential still give a contribution to the 
string tension. 

\begin{figure*}[!htb]
\begin{center}
\epsfxsize=8.0cm \epsffile{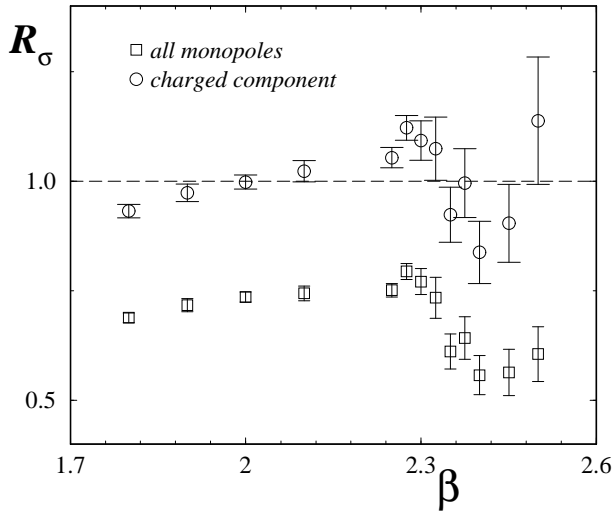} 
\end{center}
\vspace{-0.5cm}
\caption{The ratio of the temporal string tensions~\eq{Rsigma}
{\it vs.}~$\beta$.}
\label{fig:string:ratio}
\end{figure*}

It is interesting to check how the monopole density fits into the theoretical
predictions of the string tension~\eq{zero:temp:string}.
Using that predicted relation, we compare in Figure~\ref{fig:string:ratio}
%%%%%%%%%%%%%%%%
the ratio $R$ between the measured string tension $\sigma$ 
(from plane--plane correlators of Polyakov loops)
with a calculated ``theoretical'' string tension $\sigma^{\mathrm{th}}$ 
using as input the {\it measured} monopole density $\rho$:
\beqn
  R_\sigma = \frac{\sigma}{\sigma^{\mathrm{th}}}\,.
\label{Rsigma}
\eeqn
Here $\sigma^{\mathrm{th}}$ is given in accordance to 
eqs.(\ref{rho:theory},\ref{zero:temp:string}) via
\beqn
  \sigma^{\mathrm{th}} = \frac{4}{\pi} \, 
  \sqrt{\frac{\rho(\beta)}{\beta_V(\beta)}}\,,
  \label{theor:string}
\eeqn
and $\beta_V$ is defined in eq.~\eq{betaV}. 

The circles shown take into account all ``active'' monopoles, {\it i.e.} 
isolated ones and those from charged monopole clusters which 
might be thought to be responsible for the string tension. 
The ratio is close to unity indicating the fact that the charged monopoles 
provide the 
major contribution to the string tension, as expected.
Note that in the deconfinement phase the string tension is non-zero due to  
finite--size
effects discussed below. The squares in Figure~\ref{fig:string:ratio}(a) are 
related to the 
ratio~\eq{Rsigma} in which all monopole are taken into account. 
In both phases this ratio is smaller than unity: 
a neutral fraction of the monopoles bound in the small dipole pairs 
does not contribute to the string tension.

The small ``string tension'' remaining after 
passing the deconfinement transition at this finite lattice can be explained 
from the point of view of the dipole picture as follows.  Test particles 
separated by distances smaller than sizes of certain dipoles are influenced by
the constituent monopoles of those dipoles. The monopoles give contribution 
to the string tension term. On the finite lattice the maximal distance between 
the test particles is of the order of the lattice size. 
Therefore, dipoles of the same size could be responsible for the non-vanishing 
small string tension. Dipoles of these sizes may really be present in the 
deconfinement phase (with a probability decreasing with the increase 
of the lattice size). This, however, does not contradict the criterion used to 
locate the phase transition in the previous Sections since the dipoles of such 
large size are heavily suppressed.

\begin{figure*}[!htb]
\begin{center}
\epsfxsize=8.0cm \epsffile{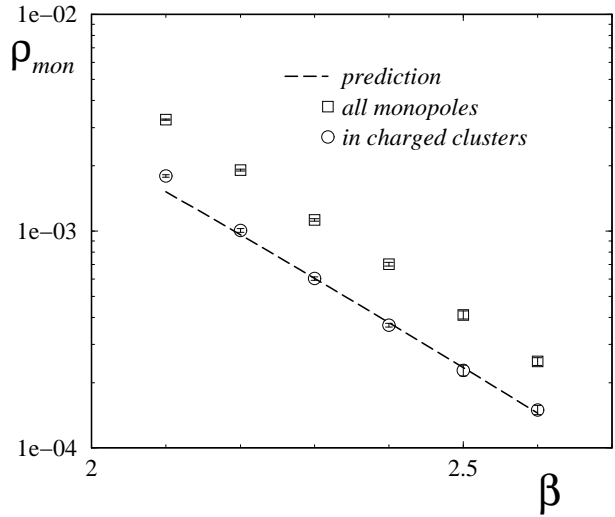} 
\end{center}
\vspace{-0.5cm}
\caption{The density of all monopoles and of monopoles in charged clusters 
{\it vs.}~$\beta$
for a $32^3$ lattice compared to  prediction~\eq{rho:theory}.}
\label{fig:mondens:zeroT}
\end{figure*}

The dipole formation due to Coulomb forces also happens at zero temperature. 
This effect
increases the monopole density compared to that in the "bindingless" world. 
To check this we
compare in Figure~\ref{fig:mondens:zeroT} the total density of 
%%%%%%%%%%%%%%%%%%%%%%%%%%
the total density of monopoles and the exclusive density of 
monopoles residing in charged clusters 
(the latter includes free monopoles and anti-monopoles)
for a $32^3$ lattice. The charged monopoles comprise around $55\%$ 
of the total monopole density. 
This ratio does not depend on the value of the coupling
constant $\beta$ indicating that the scaling behaviour 
of charged and neutral clusters is the
same. The charged fraction of the monopoles is perfectly described by
the "bindingless" formula~\eq{rho:theory} for the monopole density. 
This formula is 
incorporated implicitly into the theoretical prediction of the string 
tension~\eq{zero:temp:string} which works well according to Ref.~\cite{Stack}.
Thus only the monopoles from the charged clusters 
(including separate monopoles) contribute to the string tension
while the binding effect causes the appearance of 
a large fraction of ``inactive'' neutral clusters.

\begin{figure*}[!htb] 
\begin{center} 
\epsfxsize=7.0cm \epsffile{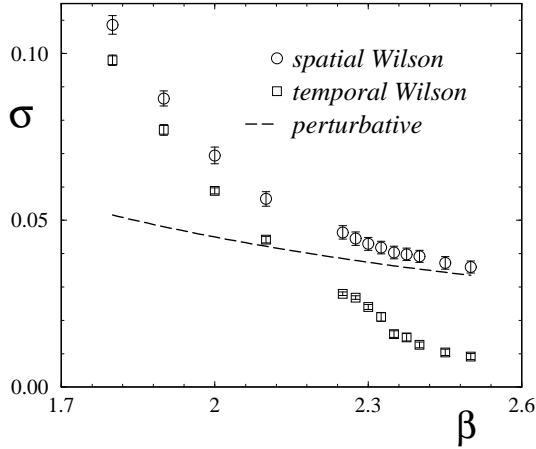}
\caption{The spatial string tension {\it vs.} $\beta$.}
\label{fig:string:spatial} 
\end{center} 
\end{figure*}

Finally, we have measured the ``spatial string tension'': the coefficient in 
front of the area term in the spatial Wilson loops. This string tension 
$\sigma_s$ 
has been obtained by means of the standard diagonal Creutz ratios. The results 
are presented in Figure~\ref{fig:string:spatial} 
%%%%%%%%%%%%%
as a function of $\beta$. 
As expected, the spatial string tension does not vanish and behaves smoothly 
across the deconfining phase transition. In contrast, we show in this Figure 
also the ``true'' temporal string tension extracted from temporal Wilson loops 
which drops down to the 
level of the finite-volume correction that we have just discussed. 

At sufficiently high temperatures the system might be treated as 
two--dimensional 
with an effective $2D$ coupling constant $\beta^{(2D)} = L_t \, \beta$. 
Moreover, since the monopole density  is low at large $\beta$ 
(in deconfinement), the model 
becomes effectively non--compact. Thus the spatial Wilson loop behaviour in 
this regime is given by the perturbative one--photon exchange. 
In two dimensions the Coulomb law provides the
linearly confining potential, $V^{(2D)}(R) = R \slash 2$, corresponding to
the spatial string tension,
\beqn	
  \sigma^{\mathrm{th}}_s(\beta) = 
  \frac{1}{2 \beta^{(2D)}} = \frac{1}{2 L_t \, \beta}\,.
\eeqn
which  is shown in Figure~\ref{fig:string:spatial} by the dashed line. The  
spatial string tension data and the curve approach each other for sufficiently
large $\beta$. However, in the confinement phase the monopoles give a 
significant contribution to the spatial string tension.

\section{Summary}

In this paper we have considered a mechanism of the finite temperature 
deconfinement phase
transition in three dimensional compact electrodynamics based on the monopole 
binding. The
considerations are similar to those given in Ref.~\cite{AgasianZarembo} for 
the continuum
theory and they incorporate features of the lattice geometry. This allows us 
to predict the
pseudocritical coupling as a function of the lattice size.

In our numerical simulations we have demonstrated that the monopoles are 
sensitive to the phase
transition despite the fact that the monopole density itself behaves smoothly 
across the
transition. The pseudocritical couplings found by the Binder cumulants of the 
density are very
close to that identified using the Polyakov loop susceptibility. 
We stress that we did not
intend to study the finite size scaling behavior
of this model.

Based on the observation to find $\beta_c$ 
in this way we have studied the monopole properties in more 
detail. We have found that both the monopole density and the string tension 
differ from the predictions based on a model which does not take into
account the monopole binding effects. However we have found numerically that 
the ratio between these two quantities derived in that model 
(given by eq.~\eq{zero:temp:string}) remains valid in the confinement phase.

We have observed that the dipole formation happens both in the confinement 
and deconfinement phases. In the deconfinement phase tightly bound  
dipoles --- which are safely identified by a cluster algorithm --- 
dominate in the vacuum. The dipoles are oriented dominantly in the temporal 
direction. These features are in agreement with general
expectations discussed in the Introduction and in Section~\ref{sec:physics}. 

At the confinement phase transition we observe mostly  clusters with two 
constituents or single monopoles and
anti--monopoles.  Decreasing further  
the temperature (or $\beta$), the monopoles become dense and form connected
clusters (on a coarser and coarser lattice) inclosing various numbers of 
monopoles and antimonopoles. The largest clusters are more and more spherical.
Whether the observed properties of the dipole gas formation survives in
the continuum limit deserves an additional study.

When the phase transition is mediated by charged objects, one could expect
that external fields will influence the phase transition.
In our case the natural question arises what will happen to the
confinement--deconfinement phase transition. 
For non--Abelian theories in $3+1$ dimensions it was recently concluded,
from a study of the expectation value of the Polyakov loop~\cite{Cea},
that confinement seems to become restored under the influence
of an external chromomagnetic field.
In an accompanying paper~\cite{paper2}
we will report on a study of our model under such external conditions,
concerning the influence of confinement and relevant properties of the 
monopole system.

\section*{Acknowledgements}

Authors are grateful to P.~van Baal, M.I.Polikarpov and J. Zaanen for 
interesting discussions.
M.N.Ch. acknowledges a support of S\"achsisches Staats\-ministerium f\"ur 
Kunst und Wissenschaft,
grant 4-7531.50-04-0361-01/16 and kind hospitality of NTZ and the 
Institute of Theoretical Physics
of Leipzig University.  
Work of M.N.Ch., was partially supported by grants RFBR 99-01230a, RFBR
01-02-17456 and CRDF award RP1-2103.  
E.-M.I. acknowledges the support by the Graduiertenkolleg
``Quantenfeldtheorie'' for a working visit to Leipzig.

\end{document}